\documentclass[varenna]{cimento}

\usepackage{graphicx}  

\title{Sewing Greenberger-Horne-Zeilinger states with a quantum zipper}

\author{Da-Wei Wang}

\institute{Institute of Quantum Science and Engineering, Texas A$\&$M University, College Station, TX 77843, USA}

\begin{document}

\maketitle

\begin{abstract}
A mechanism of a chiral spin wave rotation is introduced to systematically generate mesoscopic Greenberger-Horne-Zeilinger states.
\end{abstract}

\section{Introduction}

One of the astounding principles of quantum mechanics is that a quantum object can exist in a superposition state, \textit{e.g.}, a spin in a superposition of up and down states, $|\psi\rangle=(|\uparrow\rangle+|\downarrow\rangle)/\sqrt{2}$. Applying this principle to two quantum particles, a superposition state such as a singlet state $|\psi\rangle=(|\uparrow\downarrow\rangle-|\downarrow\uparrow\rangle)/\sqrt{2}$ allows the instantaneous determination of the quantum state of the second particle after one measures the quantum state of the first particle, as revealed by Einstein, Podolsky and Rosen in 1935 \cite{Einstein1935}. This instantaneous determination seems like a violation of a principle of the special relativity, \textit{i.e.}, nothing can travel faster than light, leading to the so called EPR paradox. The incompleteness of quantum mechanics and the existence of hidden variables were therefore suspected. 

In a seminal paper by Bell in 1964 \cite{Bell1964}, an equality was proposed to show that the statistical results of quantum mechanics is incompatible with any local hidden variable theories. Subsequent experiments overwhelmingly support the predictions of quantum mechanics. In particular, any local hidden variable theories are excluded by loophole-free experiments in 2015 \cite{Hensen2015}.

In the exploration of EPR paradox, the Greenberger-Horne-Zeilinger (GHZ) states \cite{Greenberger1990} play an important role. It concerns with a three-particle entangled state, $|\psi\rangle=(|\uparrow\uparrow\uparrow\rangle+|\downarrow\downarrow\downarrow\rangle)/\sqrt{2}$, which allows a direct contradiction between the predictions of quantum mechanics and local hidden variable theories with a single test, in contrast to the statistical nature of the Bell's inequality. GHZ states have also been found useful in the Heisenberg limit metrology \cite{Leibfried2004}. Mesoscopic GHZ states with $M$ particles $|\psi\rangle=(|\uparrow~\rangle^{\otimes M}+|\downarrow~\rangle^{\otimes M})/\sqrt{2}$ can be generated by using the M{\o}lmer-S{\o}rensen approach \cite{Molmer1999} in ion traps, with $M$ up to 14 in experiments \cite{Monz2011}. To generate GHZ states in a wide range of quantum systems, new mechanisms are highly wanted. Here we introduce a ``quantum zipper" to sew GHZ states in a systematic way.

\begin{figure}
\includegraphics[scale=0.27]{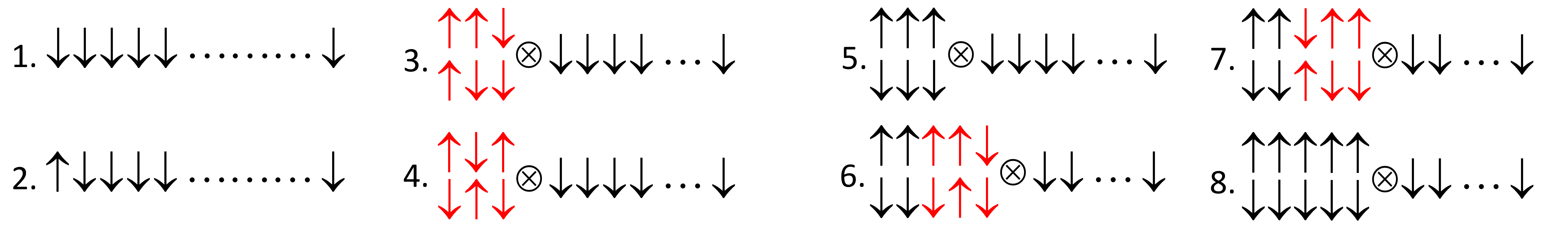}
\caption{Basic steps of a quantum zipper sewing GHZ states. 1, initialize an all-down state. 2, flip the first spin. 3, prepare the second spin in a superposition of the up and down states. 4, introduce an interaction among the red spins, which leads to chiral spin wave rotations as shown in eq.(\ref{eq1}). 5, send a $\pi$ pulse to flip the second spin. 6, send in a $\pi$ pulse to flip the fourth spin. 7, introduce the same interaction as in step 4 among the red spins until achieving the state shown in this step. 8, send a $\pi$ pulse to flip the third spin. Repeat the steps 6-8 until zipping all spins in the GHZ state, each time adding two spins to the GHZ chain. For a GHZ state with $2n+1$ spins, we need in total $2n$ $\pi$ pulses and one $\pi/2$ pulse.}
\label{scheme}
\end{figure}

The basic process for generating GHZ states is shown in fig. \ref{scheme}. We start with an all-down state and send a $\pi$ pulse to the first spin to flip it up. Then we send a $\pi/2$ pulse to prepare the second spin in a superposition of the up and down states. The following step 4 is crucial. We introduce a special interaction between the first three spins, which undergo opposite chiral spin wave rotations for $|\uparrow\downarrow\downarrow\rangle$ and $|\downarrow\uparrow\uparrow\rangle$ states,

\begin{eqnletter}
\label{eq1}
|\uparrow\downarrow\downarrow\rangle \rightarrow |\downarrow\uparrow\downarrow\rangle \rightarrow |\downarrow\downarrow\uparrow\rangle \rightarrow |\uparrow\downarrow\downarrow\rangle,    \label{ea}\\
|\downarrow\uparrow\uparrow\rangle \rightarrow |\uparrow\uparrow\downarrow\rangle \rightarrow |\uparrow\downarrow\uparrow\rangle \rightarrow |\downarrow\uparrow\uparrow\rangle.      \label{eb}
\end{eqnletter}
The spin states move to the right for the states containing one up spin while they move to the left for the states containing two up spins. The step 4 corresponds to the rotations $|\uparrow\uparrow\downarrow\rangle \rightarrow |\uparrow\downarrow\uparrow\rangle$ and $|\uparrow\downarrow\downarrow\rangle \rightarrow |\downarrow\uparrow\downarrow\rangle$ in eq.(\ref{eq1}). We then send a $\pi$ pulse to the second spin and prepare the first three spins in a GHZ state, as shown in the step 5. Following and repeating the steps 6-8 (see fig.\ref{scheme} caption), we can zip the following spins into a GHZ state.

\section{Mechanism}

In a previous work, we studied chiral photon rotations among three cavities by generating a synthetic magnetic field \cite{Wang2016}. The process in eq.(\ref{eq1}) can be realized in a similar way by the following Hamiltonian,
\begin{equation}
H=i\hbar\kappa\sum\limits_{j=1}^{3} \sigma^+_{j+1} \sigma^-_j+h.c.,
\label{Hspin}
\end{equation}
where $\sigma^+_j$ and $\sigma^-_j$ are the raising and lowering operators for the $j$th spin and the summation over $j$ is cyclic. The Hamiltonian in eq.(\ref{Hspin}) commutes with $\sum \sigma^z_j$ and the number of up spins is conserved. We first investigate the dynamics in the subspace expanded by $|\uparrow\downarrow\downarrow\rangle$, $|\downarrow\uparrow\downarrow\rangle$ and $|\downarrow\downarrow\uparrow\rangle$, in which the eigen frequencies are $\lambda_1=0$, $\lambda_2=\sqrt{3}\kappa$ and $\lambda_3=-\sqrt{3}\kappa$. The corresponding eigenstates are
\begin{eqnletter}
|\psi_1\rangle &=& \frac{1}{\sqrt{3}}(|\uparrow\downarrow\downarrow\rangle + |\downarrow\uparrow\downarrow\rangle + |\downarrow\downarrow\uparrow\rangle),\\
|\psi_2\rangle &=& \frac{1}{\sqrt{3}}(|\uparrow\downarrow\downarrow\rangle + e^{i2\pi/3}|\downarrow\uparrow\downarrow\rangle + e^{i4\pi/3}|\downarrow\downarrow\uparrow\rangle),\\
|\psi_3\rangle &=& \frac{1}{\sqrt{3}}(|\uparrow\downarrow\downarrow\rangle + e^{i4\pi/3}|\downarrow\uparrow\downarrow\rangle + e^{i2\pi/3}|\downarrow\downarrow\uparrow\rangle).
\end{eqnletter}
The evolution of the initial state $|\Psi(0)\rangle=|\uparrow\downarrow\downarrow\rangle=(|\psi_1\rangle+|\psi_2\rangle+|\psi_3\rangle)/\sqrt{3}$ is
\begin{eqnarray}
\label{wf}
|\Psi(t)\rangle &=& \frac{1}{\sqrt{3}}\sum\limits_{j=1}^{3}e^{-i\lambda_j t}|\psi_j\rangle
=\frac{1}{3}[(1+2\cos(\sqrt{3}\kappa t))|\uparrow\downarrow\downarrow\rangle\\
\nonumber
&+&(1+2\cos(\sqrt{3}\kappa t-2\pi/3))|\downarrow\uparrow\downarrow\rangle
+(1+2\cos(\sqrt{3}\kappa t+2\pi/3))|\downarrow\downarrow\uparrow\rangle].
\end{eqnarray}
It is clear that at time $t=T\equiv 2\pi/(3\sqrt{3}\kappa)$, $|\Psi(T)\rangle=|\downarrow\uparrow\downarrow\rangle$, and at time $t=2T$, $|\Psi(2T)\rangle=|\downarrow\downarrow\uparrow\rangle$. We obtain the chiral spin wave rotation in eq.~(\ref{eq1}$a$).

We can follow the same procedure to calculate the dynamics in the subspace with two up spins, $|\downarrow\uparrow\uparrow\rangle$, $|\uparrow\downarrow\uparrow\rangle$ and $|\uparrow\uparrow\downarrow\rangle$. The spin states move to the left, as shown in eq.~(\ref{eq1}$b$). To understand this surprising result, we try to know how $|\downarrow\uparrow\uparrow\rangle$ evolves based on the knowledge that the state $|\uparrow\downarrow\downarrow\rangle$ rotates to the right. By reversing our definition of up and down, the state $|\downarrow\uparrow\uparrow\rangle$ becomes $|\uparrow\downarrow\downarrow\rangle$ in the new basis. To express the Hamiltonian in this upside down basis, we need to make the replacement $\sigma^+_j\rightarrow\sigma^-_j$ and $\sigma^-_j\rightarrow\sigma^+_j$, which results in $H\rightarrow-H$. The state evolves backward in time, \textit{i.e.}, the spin wave moves to the left.

\section{Implementation}

The key feature of the Hamiltonian in eq.(\ref{Hspin}) is the imaginary interaction strength between the spins. Previously we proposed the synthetic magnetic field for photons by oscillating the frequencies of three cavities that are coupled to the same spin \cite{Wang2016}. Here we consider three spins with oscillating frequencies coupled to the same cavity mode. Let us first consider two spins with frequencies being modulated with different phases. The interaction Hamiltonian can be written as
\begin{equation}
H_I=\hbar g a^\dagger [\sigma^-_1 e^{if\cos(\nu_d t+\phi_1)}+\sigma^-_2 e^{if\cos(\nu_d t+\phi_2)}]+H.c.,
\end{equation}
where $a$ is the annihilation operator of the cavity, $f$, $\nu_d$ and $\phi_j$ are the modulation amplitude, frequency and phase for the $j$th spin. We assume the central frequencies of the spins are the same as that of the cavity. Since $e^{if\cos(\nu_d t+\phi_j)}=\sum_{n=-\infty}^{\infty}i^{n}J_{n}(f)e^{in(\nu_dt+\phi_j)}$ where $J_{n}(f)$ is the $n$th order Bessel function of the first kind, we can expand the interaction Hamiltonian $H_I=\sum_{n} h_{n}e^{in\nu_{d}t}$, where
\begin{equation}
h_n=i^nJ_n(f)\hbar g a^\dagger [\sigma^-_1 e^{in\phi_1}+\sigma^-_2 e^{in\phi_2}]
+i^nJ_n(-f)\hbar g [\sigma^+_1 e^{in\phi_1}+\sigma^+_2 e^{in\phi_2}]a,
\end{equation}
The effective Hamiltonian is \cite{Wang2016},
\begin{equation}
H_{e}=h_0+\sum\limits_{n=1}^{\infty}\frac{1}{n\hbar\nu_d}[h_n,h_{-n}]\\
=h_0+i\frac{\hbar g^2}{\nu_d}\eta(\sigma_2^+\sigma_1^--\sigma_1^+\sigma_2^-).
\end{equation}
where $\eta=2\sum_{n=1}^{\infty}J^2_n(f)\sin[n(\phi_2-\phi_1)]/n$. When $f=2.4$, $J_0(f)=0$ and $h_0=0$, we obtain the imaginary interaction strength between the two spins. We introduce the third spin with the same modulation amplitude $f$ and frequency $\nu_d$ but a different phase $\phi_3$, and we set $\phi_j=2j\pi/3$. We obtain $H_{e}=H$ in eq. (\ref{Hspin}) with $\kappa=g^2\eta/\nu_d$ and $\eta=0.307$.

In the experiments, the transition frequencies of the spins can be modulated by the dynamic Stark shift of a detuned light field with modulated intensities. The spins can be atoms coupled to a one-dimensional waveguide that can mediate long range interactions between atoms \cite{Chang2013a}.

\end{document}